\documentclass[fleqn,12pt,twoside]{article}
\usepackage[headings]{photon_hbt_star}

\readRCS
$Id: espcrc1.tex,v 1.2 2004/02/24 11:22:11 spepping Exp $
\usepackage{graphicx}
\usepackage[figuresright]{rotating}

\newcommand{\AmS}{{\protect\the\textfont2
  A\kern-.1667em\lower.5ex\hbox{M}\kern-.125emS}}

\title{Preliminary Results on Direct Photon-Photon HBT Measurements in $\sqrt{s_{\rm{NN}}}$ = 62.4 GeV and 200 GeV Au+Au Collisions at RHIC}

\author{Debasish Das \address[VECC]{ Variable Energy Cyclotron Centre, 
        1/AF  Bidhan Nagar, \\ 
        Kolkata - 700064, INDIA. }\thanks{ ddas@veccal.ernet.in },
        Guoji Lin \address[YALE]{ Yale University, Physics Department\\
         P.O 208120, New Haven \\
         Conneticut 06520, USA.}\thanks{ guoji.lin@yale.edu },
        Subhasis Chattopadhyay \addressmark[VECC],
        Alexei Chikanian \addressmark[YALE],
        Evan Finch \addressmark[YALE],
        Tapan K Nayak \addressmark[VECC],
        Sergey Y Panitkin \address[BNL]{ Physics Department,\\
        Brookhaven National Laboratory\\
        P.O Box 5000, Upton, NY 11973, USA.},
        Jack Sandweiss \addressmark[YALE],
        Alexandre Alarcon Suaide \address[BNL]{ Instituto de Fisica da Universidade de Sao Paulo,\\
        P.O. Box 66318, Sao Paulo 05314-970, BRAZIL. },
        Haibin Zhang \addressmark[BNL]{} 
        for the STAR Collaboration.}
       
\runtitle{Direct Photon-Photon HBT Measurements at RHIC}
\runauthor{D.Das, G.Lin  et al. (STAR Collaboration).}

\begin{document}

% typeset front matter
\maketitle

\begin{abstract}
We present the preliminary results on direct photon interferometry measurements
in Au+Au collisions at  $\sqrt{s_{\rm{NN}}}$ = 62.4 and 200 GeV using the STAR ( Solenoidal
Tracker at RHIC ) detector. Photons are reconstructed via $e^+/e^-$ conversions in STAR Time
Projection Chamber(TPC) and energy deposited by photons in STAR Barrel Electromagnetic Calorimeter(BEMC). The two-photon correlations are measured using (1) both photons measured using BEMC; (2) one photon from conversions and the other measured with BEMC. Both the methodologies and the possible constraints in the correlation function measurements are discussed.
\end{abstract}

\section{Introduction}

          The information about the space-time structure of the emitting source created in elementary particle and heavy-ion collisions from the measured particle momenta can be extracted by the method of two-particle intensity interferometry techniques [Hanbury Brown - Twiss (HBT)]\cite{nature,boal}. Experimentally, the two-particle correlation function (normalized to unity at large $Q_{inv}$) is obtained from the ratio

\begin{equation}
C_{2}(Q_{inv}) = A(Q_{inv})/B(Q_{inv}) ,
\end{equation}where A($Q_{inv}$) is the pair distribution in invariant momentum for particle-pairs from the same event, and B($Q_{inv}$) is the corresponding distribution for pairs of particles taken from different events \cite{heinz}.

         Historically most of the HBT measurements in heavy-ion experiments have been done with the pions \cite{star130,star200} and also have been extended to kaons, protons and other heavier particles. Hadron correlations reflect the properties of the hadronic source, i.e. size of the system at the freeze-out time. The direct photons, which are emitted during all the stages of the collision, serve as a deep probe of the hot and dense matter. Hence direct photon HBT correlations can provide the system sizes at all stages of heavy-ion collisions \cite{kapus,kapus1,timmer}.  

          Direct photons emitted from the early hot phase of the relativistic heavy ion collisions and their HBT correlations serve as important signatures of the quark gluon plasma and its properties. Due to their electromagnetic nature of interaction, the photons have a clear advantage in such studies \cite{alam} as they weakly interact with the system and are free from Coulomb interactions, which one needs to correct for hadronic HBT measurements. 

      The photon-photon Bose-Einstein interferometry of the direct photons provides information about the various stages of heavy-ion collisions \cite{bass}. However, it is difficult to extract the small yield of direct photons due to the large background of photons produced by electromagnetic decay of the hadrons (especially $\pi^{0}$s and $\eta$). It has been proposed that one can obtain the direct photon HBT signal using all produced photon HBT signal using all produced photons \cite{peres}. Direct photon HBT correlations were observed at SPS energies \cite{peres1}. 

      In this paper we present the preliminary results on direct photon interferometry at  $\sqrt{s_{NN}}$ = 62.4 GeV and 200 GeV Au+Au collisions in the Solenoidal Tracker at RHIC (STAR) experiment at the Relativistic Heavy Ion Collider (RHIC) facility in Brookhaven National Laboratory, USA. We have analysed minimum bias events for both the Au+Au collisions datasets.

      Photons are reconstructed using two techniques: (a) energy deposited by photons in the STAR Barrel Electromagnetic Calorimeter (BEMC) and (b) from conversions in STAR Time Projection Chamber (TPC) \cite{tpcnim}. The photon correlation functions $C_{2}$($Q_{inv}$) are calculated using two ways: (1) both photons measured using BEMC and (2) one photon from conversions and the other measured with BEMC.  The photon identification efficiency in TPC is low and the granularity of BEMC tower hinders the observation of two very close BEMC photons. The energy threshold of the photons reconstructed from BEMC is relatively large. So the combination of BEMC-TPC photon pairs can help to provide a compromise between these individual photon detection limitations.
\section{The STAR detector and experimental setup}
 
     The STAR detector is one of the two large-scale 
ongoing experiments at RHIC. The present analysis is  
based on the Au+Au collisions measured by STAR 
detector in Run-IV. In the following, we describe the 
detectors that are relevant to the present analysis. The  
Time Projection Chamber (TPC) is used in STAR as the 
primary tracking device. The TPC records the particle  
tracks, measures their momenta and hence provides the 
particle identification by measuring their ionization 
energy loss ($dE/dx$). Particles are identified for a 
momentum range of 100 MeV/c to greater than 1 GeV/c 
and the measured momenta ranges from 100 MeV/c to  
30 GeV/c. TPC has an acceptance of $|{\eta}|$ $<$ 1.8 with full 
azimuthal coverage. Photon momentum reconstruction 
was done by the measured momenta of $e^+/e^-$  pairs 
produced from conversions.

     The Barrel Electro-Magnetic Calorimeter (BEMC) 
is the main detector in STAR for photon measurements 
\cite{emcnim}. For the RHIC Run-IV, 50$\%$ of the BEMC was 
installed and operational, covering an acceptance of 
0$<$$\eta$$<$1 and full azimuth. The BEMC is a lead-
scintillator sampling electromagnetic calorimeter with 
equal volumes of lead and scintillator. The calorimeter 
has a depth of 20 radiation lengths at $\eta$=0 and an inner 
radius of   220 cm.

    The BEMC includes a total of 120 calorimeter 
modules, each subtending an angle of $6^0$ in  $\phi$ direction 
($\sim$ 0.1 radian) and 1.0 unit in  $\eta$ direction. These 
calorimeter modules are mounted 60 in $\phi$ and 2 in $\eta$. 
Each module is divided into 40 towers with granularity 
($\Delta\eta$,$\Delta\phi$) = (0.05, 0.05). Two layers of gaseous Shower 
Maximum Detectors (SMDs) with two-dimensional 
readout are located at 5$X0$ inside the calorimeter 
module.

     The SMD is a wire proportional counter - strip 
readout detector using gas amplification. While the 
BEMC towers provide precise energy measurements for 
isolated electromagnetic showers, the high spatial 
resolution of ($\Delta\eta$,$\Delta\phi$) = (0.007, 0.007) provided by the 
SMDs is essential for  identification, $\pi^0$ reconstruction 
and electron identification. The electromagnetic energy 
resolution of the calorimeter is $\delta$E/E $\sim$ 16$\%$ / $\sqrt{E}$ (GeV).

     The STAR trigger detectors are the CTB and ZDCs 
but separate trigger settings are used for the data 
analysis of  Au+Au collisions at $\sqrt{s_{\rm{NN}}}$ = 62.4 GeV and 200 GeV. 
  
\section{Correlation analysis of BEMC Photons at  $\sqrt{s_{\rm{NN}}}$ = 62.4 GeV}

     For the present analysis of two photon correlation functions 
we have used 1.8 M minimum-bias events of Au+Au collisions at $\sqrt{s_{\rm{NN}}}$ = 62.4 GeV 
after the event quality cuts. 
The event quality cuts include (1) the selection of events with a collision 
vertex with $\pm$ 30cm measured along the beam axis from 
the center of TPC, and (2) the rejection of corrupted 
events with high spurious energy measurements. The 
details of energy reconstruction and measurements in 
EMC can be found in \cite{emcet}. The photon-like clusters 
in the calorimeter are taken in this analysis which do not 
have a TPC track pointing at them within the proximity 
of single tower ($\Delta\eta$,$\Delta\phi$) = (0.05,0.05).

     The two-photon correlation function, defined 
in Eq. (1) was calculated as the ratio of the distribution 
of photon pair invariant relative momenta, $Q_{inv}$.

     During the study of correlations at small $Q_{inv}$, we 
need to understand the effects produced by splitting of a 
single electromagnetic shower into multiple showers, or 
merging of nearby showers into a single cluster. From 
the studies of opening angle between two photon 
clusters measured in EMC it was found that the 
minimum angle that reduces cluster merging is $\sim$0.02 
radian. 

      The effects of cluster splitting have been studied 
with opening angle and cluster energy cuts. Since the 
splitting photons are spatially close we rejected all 
photon clusters from the same tower.  

\begin{figure}[h]
\begin{center}
\includegraphics[width=30pc]{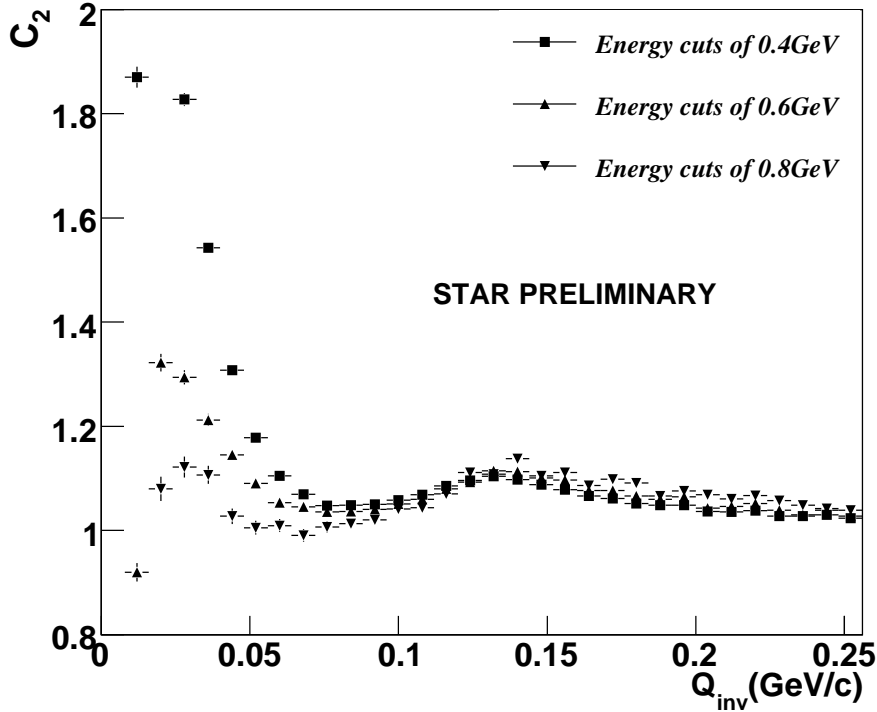}\hspace{2pc}
\end{center}
\begin{center}
\begin{minipage}[b]{36pc}\caption{\label{vecc}{\bf{Preliminary}} two-photon correlation function for neutral BEMC clusters with different energy cuts. Here BEMC clusters are selected from different adjacent towers. The region below the $\pi^0$  peak at $Q_{inv}$ $\approx$ $m_{\pi^0}$ is where the interference effects become significant.}
\end{minipage}
\end{center}
\end{figure}

      Further studies on two-photon correlation function 
with photons taken from different towers along with 
energy cuts are also carried out. The two-photon 
correlation functions with pairs having energy 
specifically above 0.4, 0.6 and 0.8 GeV are shown in 
 Fig.~\ref{vecc}. 
 
      It can be seen from Fig.~\ref{vecc} that photon energy cut 
 helps to reduce large enhancement below 0.1 GeV/c. 
 This enhancement is related with the lepton pairs from 
 the photon conversions in outer field cage of the TPC. 
 We also see a clearly visible peak at $Q_{inv}$ $\approx$ $m_{\pi^0}$
 due to  the neutral pion decays. The energy cuts applied on the 
 photon pairs reduce splitting but such cuts also require 
 optimization as the correlation strength for direct 
 photons is much less than the decayed ones. At present 
 we are studying how judiciously we can disentangle the
 apparatus effects, which mimic direct photon Bose-
 Einstein correlations. 
 
%\begin{figure}[h]
%\begin{minipage}[t]{80mm}
%\framebox[79mm]{\rule[-26mm]{0mm}{52mm}}
%\includegraphics[width=30pc]{Figure1.eps}
%\caption\label{Good sharp prints should be used and not (distorted) photocopies.}
%\label{fig:largenenough}
%\end{minipage}
%

       We have also tested other properties of cluster 
splitting, like relative energy difference. In the case of 
erroneous cluster splitting one part is assigned more 
energy than the other. So an effective cut to remove 
such pairs can help us to reduce splitting. We have 
studied the effects of two-particle energy asymmetry 
(alpha cut) $|E_{1} - E_{2}| / |E_{1} + E_{2}|$ on correlations functions 
and found that in the case of STAR BEMC this cut is not effective in the removal of cluster 
splitting.

%\hspace{\fill}
%\begin{minipage}[t]{75mm}
%\framebox[74mm]{\rule[-26mm]{0mm}{52mm}}
%\caption{Remember to keep details clear and large enough to
%withstand a 20--25\% reduction.}
%\label{fig:toosmall}
%\end{minipage}
%\end{figure}

\section{Correlation analysis of TPC-BEMC Photons at  $\sqrt{s_{\rm{NN}}}$ = 200 GeV}

         In this analysis we have taken 5 M year-IV Au+Au minimum-bias events at $\sqrt{s_{\rm{NN}}}$ = 200 GeV after event quality cuts. We keep events satisfying the standard minimum-bias trigger with a collision vertex of $\pm$30cm along the beam-axis from the center of TPC. 
      
          The TPC photons are reconstructed via $e^+/e^-$ conversions \cite{ianjohn}. For the present analysis of two photon correlation functions, one photon is taken from conversions and the other measured with BEMC. Electron and positron tracks are selected from particle identification by energy loss in the TPC. The number of space points of each track should be greater than 12. The ratio of the number of space points to the expected maximum number of space points for that particular track is required to be greater than 0.55 to largely suppress track splitting.

%\begin{center}
\begin{figure}[h]
\begin{center}
\includegraphics[width=30pc]{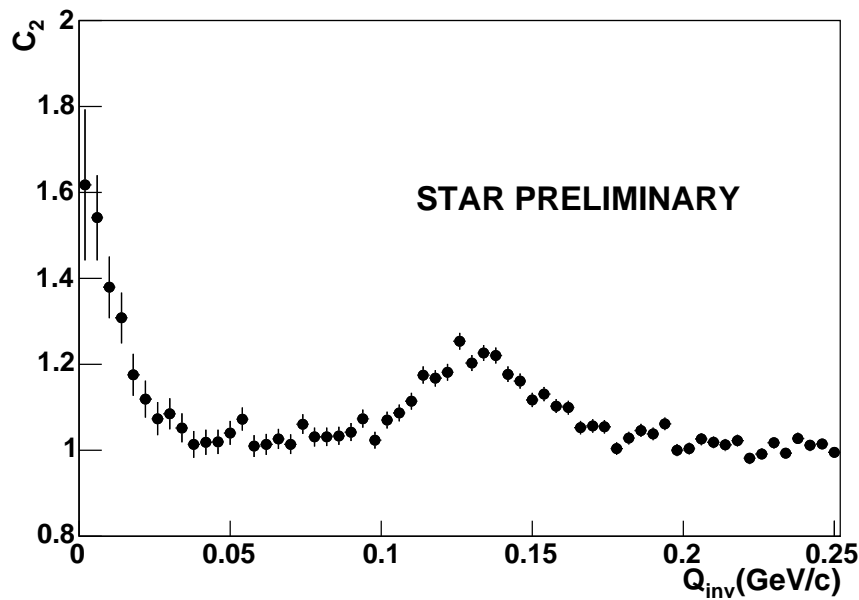}\hspace{2pc}
\end{center}
\begin{center}
\begin{minipage}[b]{36pc}\caption{{\label{yale}\bf{Preliminary}} two-photon correlation function
 using the photon pairs from BEMC and TPC.}
\end{minipage}
\end{center}
\end{figure}
%\end{center}

          A converting primary photon is emitted from the primary vertex. It converts into $e^+/e^-$ pair in the space with a small opening angle between two daughter tracks. As a result, we apply some geometrical and quality cuts to each $e^+/e^-$ pair to remove those random combinations. The cuts include: (1) the meeting of these two helical tracks at a conversion point inside TPC or the inner cage if extended backward, (2) the small opening angle between the momenta of two tracks at the conversion point, (3) the momentum of the reconstructed photon, i.e. the sum of two tracks' momenta at conversion point, should point back towards the primary vertex, and (4) very small invariant mass.

         The BEMC photons are measured from the energy deposited in the electromagnetic calorimeter. The BEMC towers are used to measure the energy of photons. The spatial position of the photons in the azimuthal and pseudo-rapidity directions are provided by the SMDs. Charged particles are removed by rejecting the BEMC clusters associated with any TPC charged tracks within ($\Delta\eta$,$\Delta\phi$)=(0.05, 0.05). We only use BEMC photons with energy greater than 0.8 GeV.

         The correlation function shown in Fig.~\ref{yale} uses one 
photon from BEMC while the other photon reconstructed from TPC. 
There is a big unknown peak at small $Q_{inv}$  region, with a magnitude 
greater than the theoretical limit of the direct photon HBT signal. This 
peak is insensitive to the geometrical and quality cuts of 
photons.  The study of the opening angle between photon 
pairs contributing to the low $Q_{inv}$ region reveals that 
there might be some angular correlations between 
BEMC and TPC photons. Removing TPC photons with 
energy greater than 1 GeV substantially reduces the 
peak. But we don't see a similar behavior of the 
correlation function when removing high-energy BEMC 
photons. Finally, in our simulation we observed a 
residual correlation from $\pi^0$ HBT correlation with a 
magnitude of a few percent, i.e. larger than the expected 
HBT signal, depending on the transverse-momentum 
range. This effect is still under further study.

\section{Conclusions}

     The current status of two-photon correlation functions  
using two independent methodologies, are measured for 
the first time at RHIC. We have found an interesting 
correlation at low $Q_{inv}$ and the challenge now remain to 
understand the apparatus effects and physics reasons 
behind such behaviour.


\begin{thebibliography}{9}
\bibitem{nature} R.Hanbury-Brown and R.Q.Twiss, Nature 178, 1046(1956).
\bibitem{boal} D.Boal, C.K.Gelbke and B.Jennings,Rev. Mod. Phys 62, 553(1990).
\bibitem{heinz} U.Heinz and B.V.Jacak, Ann. Rev. Nucl. Sci. 49, 529(1999). 
\bibitem{star130} C.Adler et al.(STAR Collaboration) Phys. Rev. Lett. 87, 082301(2001).
\bibitem{star200} J.Adams et al.(STAR Collaboration) Phys. Rev. C 71, 044906(2005).
\bibitem{kapus} D.K.Srivastava and J.I.Kapusta, Phys. Rev. C 48, 1335(1993). 
\bibitem{kapus1} D.K.Srivastava and J.I.Kapusta, Phys. Rev. C 50, 505(1993). 
\bibitem{timmer} A.Timmermann, M.Plumer, L.Razumov and R.M.Weiner,Phys.Rev.C 50, 3060(1994).
\bibitem{alam} J.Alam, B.Mohanty, P.Roy, S.Sarkar and B.Sinha, Phys. Rev. C 67, 054902(2003). 
\bibitem{bass} S.A.Bass, B.Muller and D.K.Srivastava, Phys. Rev. Lett 93, 162301(2004).
\bibitem{peres} D.Peressounko, Phys. Rev. C 67,014905(2003).
\bibitem{peres1} M.M.Aggarwal et al.(WA98 Collaboration) Phys. Rev. Lett. 93, 022301(2004).
\bibitem{tpcnim} M.Anderson et al., Nucl. Intrum. Meth. A 499, 659(2003).
\bibitem{emcnim} M.Beddo et al., Nucl. Intrum. Meth. A 499, 725(2003).
\bibitem{emcet} J.Adams et al.(STAR Collaboration) Phys. Rev. C 70, 054907(2004).
\bibitem{ianjohn} J.Adams et al.(STAR Collaboration) Phys. Rev. C 70, 044902(2004).

\end{thebibliography}
\end{document}